\documentclass[twoside,a4paper,11pt]{sea10}
\usepackage{graphicx}
\usepackage{hyperref}
\usepackage{movie15}
\begin{document}
\pagenumbering{arabic}
\pagestyle{myheadings}
\thispagestyle{empty}
{\flushleft\includegraphics[width=\textwidth,bb=58 650 590 680]{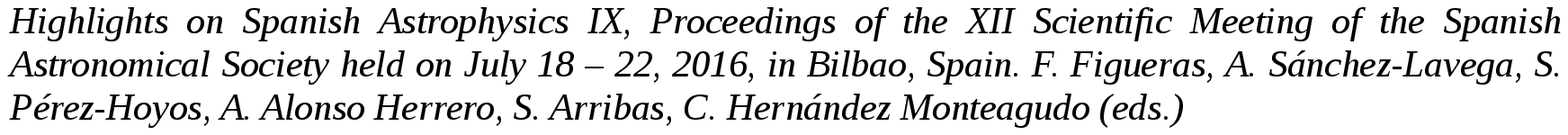}}
\vspace*{0.2cm}
\begin{flushleft}
{\bf {\LARGE
%
Uncovering star formation feedback and magnetism in galaxies with radio continuum surveys
%
}\\
\vspace*{1cm}
%
Fatemeh S. Tabatabaei$^{1,2,3}$
%
}\\
\vspace*{0.5cm}
%
$^{1}$
Instituto de Astrof\'isica de Canarias,  V\'ia L\'actea S/N, E-38205 La Laguna, Spain\\
$^{2}$
Departamento de Astrof\'isica, Universidad de La Laguna, E-38206 La Laguna, Spain\\
$^{3}$
Max-Planck-Institut f\"ur Astronomie, K\"onigstuhl 17, 69117 Heidelberg, Germany
%
\end{flushleft}
%
\markboth{
Star formation feedback and magnetism 
}{ 
%
F.\,S. Tabatabaei 
%
}
\thispagestyle{empty}
\vspace*{0.4cm}
\begin{minipage}[l]{0.09\textwidth}
\ 
\end{minipage}
\begin{minipage}[r]{0.9\textwidth}
\vspace{1cm}
\section*{Abstract}{\small
%
 Recent  studies  show  the  importance  of  the  star  formation  feedback  in  changing  the energetic  and  structure  of  galaxies.  Dissecting  the   physics  of  the  feedback  is  hence crucial  to  understand  the  evolution  of  galaxies.  Full  polarization  radio  continuum surveys can be ideally performed to trace not only star formation but also the energetic components  of  the  interstellar  medium  (ISM), the magnetic fields and cosmic ray electrons.  Using the  SKA  precursors,  we investigate the  effect  of the massive  star formation on the ISM energy balance in nearby galaxies. Our multi-scale and multi-frequency surveys show that cosmic rays are injected in star forming regions and lose energy propagating away from their birth place. Due to the star formation feedback, cosmic ray electron population becomes younger and more energetic. Star formation also amplifies the turbulent magnetic field inserting a high pressure which is important in energy balance in the ISM and structure formation in the host galaxy. 
%
\normalsize}
\end{minipage}
%
%
%
\section{Introduction \label{intro}}
To understand the evolution and appearance of galaxies it is crucial to study the ISM/star formation interplay in galaxies. Most of our information about the ISM relates to its massive component, the gas and its various phases. Its interplay with  star formation  can be addressed through the famous Kennicutt-Schmidt relation (Schmidt 1959, Kennicutt 1998) between the rate of star formation and the gas density. The Herschel and Spitzer space telescopes have made major breakthroughs in this area, by mapping the emission from dust and the gas content of galaxies.  However, not much is known about the connection of star formation and the most energetic ISM components, the cosmic rays and magnetic fields, and their role in structure formation in galaxies. As an extinction-free tracer of star formation, the radio continuum emission provides crucial complementary information about magnetic fields and cosmic ray electrons (CREs). New developments in radio correlators and polarimeters will revolutionize radio astronomy in the next years. Radio continuum observations with the forthcoming SKA and its pathfinders such as VLA, GMRT, and LOFAR are paramount to study the production and propagation of CREs, the star formation--ISM interplay, and the role of magnetic fields in structure formation and energy balance in galaxies.\\
In this proceeding, I present parts of our studies performed on resolved scales as well as globally in nearby galaxies as ideal laboratories for studying astrophysical processes that regulate the evolution of galaxies, among them the interplay between the star formation and the ISM.

\section{Resolved nonthermal emission from galaxies} 
The radio continuum emission from galaxies observed in cm-waves is mainly due to two different radiation mechanisms, the nonthermal synchrotron emission and the thermal free-free emission. The  nonthermal emission depends on the magnetic fields and CRE number density.  Therefore, separating the thermal and nonthermal emission is key to understand the sources of CREs and their distribution in galaxies and in estimates of the magnetic field strength. The classical method requires accurate absolute measurements of the brightness temperature at high and low radio frequencies and needs a prior knowledge of the nonthermal spectral index ($\alpha_n$, S$_{\nu}\sim \nu^{-\alpha_n}$).  Propagating and interacting with matter and energy in the ISM, CREs cool and their energy spectrum changes depending on the environment. Hence, a fixed $\alpha_n$  is not a correct assumption when obtaining the distributions of the nonthermal emission and  the CREs/magnetic fields across galaxies. For M33, Tabatabaei et al. (2007) developed a thermal/nonthermal separation method based on a thermal radio template (TRT) leading to the first map of the pure nonthermal spectral index in a galaxy. In star forming regions, the nonthermal spectrum is flatter ($\alpha_n \,\sim$\,0.6, typical spectral index of supernova remnants) than in the inter-arm regions and outer parts of M33 ($\alpha_n \,\sim$\,1.0, typical spectral index of synchrotron/inverse Compton loss), in agreement with the energy loss theory of CREs propagating from their places of birth (Fig.~1). The mean $\alpha_n$ across M33 is 0.95. Separating the thermal and nonthermal emission in other nearby galaxy with higher SFR, NGC~6946,  a flatter  spectrum is obtained ($\alpha_n\sim 0.8$, Tabatabaei et al. 2013). If this is due to the star formation feedback and if a general dependency exists in galaxy samples have been impossible to address based on the classical separation method. This is investigated in detail in our global study (see Sect.~4).

\begin{figure}
\center
\includegraphics[scale=0.3]{m33-ntspect.ps} ~
\includegraphics[scale=0.35]{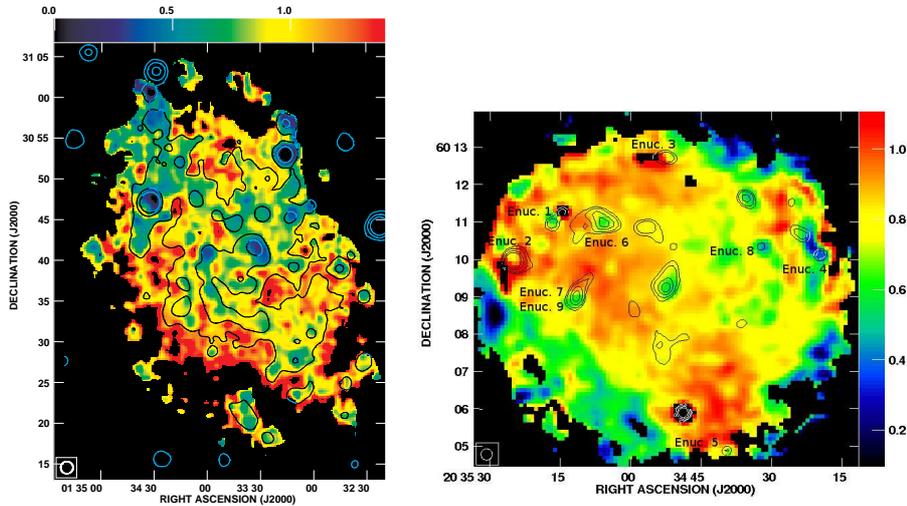} ~
\caption{{\it Left}: M33- The nonthermal spectral index ($\alpha_{n}$) map obtained from the nonthermal radio fluxes at 3.6 and 20\,cm. The spiral arms and giant H{\sc ii} regions are indicated by contours of the total radio emission at 3.6\,cm superimposed. {\it Right}: NGC~6946- $\alpha_{n}$ mapped using the same method as in M33. The 9 giant H{\sc ii} regions are indicated. The bars show the $\alpha_{n}$ values. }
\end{figure}

\section{Distribution of the magnetic field strength in galaxies}
The equipartition magnetic field strength  depends on the nonthermal intensity and spectral index $\alpha_n$ (Beck \& Krause 2005) and hence its local variation can be investigated using the maps of the pure nonthermal emission. The TRT separation technique has allowed mapping the magnetic field strength in M33 (Tabatabaei et al. 2008) and NGC~6946 (Tabatabaei et al. 2013). In both galaxies, the magnetic field strength is correlated with the SFR surface density (Fig.~2). This correlation is more likely due to the turbulent component of the magnetic field, as the SFR surface density has a tight correlation with the turbulent field. This shows an enhancement of the turbulent magnetic field due to star formation feedback, such as supernova and strong shocks. As a possible mechanism, small-scale magnetic fields could be amplified in star forming regions by a turbulent dynamo mechanism where kinetic energy converts to magnetic energy (e.g., Beck \& Hoernes 1996, Gressel et al. 2008). \\
We note that there is no correlation between the large-scale, ordered magnetic field and SFR surface density in galaxies, as in the case of M33 and NGC~6946 (Fig.~2).  Studying a sample of non-interacting/non-cluster galaxies, Tabatabaei et al. (2016) found a correlation between the large-scale magnetic field strength and the rotation speed of galaxies (Fig.~3) which shows the effect of the gas dynamics in ordering the magnetic fields in galaxies. 

\begin{figure}
\center
\includegraphics[scale=0.6]{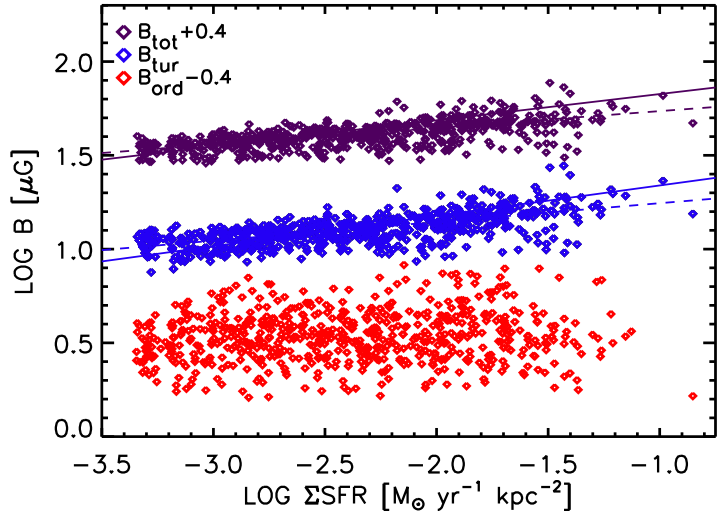}
\caption{\small Resolved magnetic field strength vs. the star formation rate in NGC 6946 (Tabatabaei et al. 2013a). Blue  symbols show the turbulent magnetic field, red symbols the ordered magnetic field (shifted by -0.4 along the Y axis), and purple symbols the total magnetic field (shifted by +0.4 along the Y axis).  The ordered magnetic field does not correlate with the SFR. }
\end{figure}

\section{Radio SEDs of galaxies}
Studying the spectral energy distribution (SED) provides significant information on the origin, energetics, and physics of the electromagnetic radiation in general. We have performed a coherent multi-band survey of the mid-radio (1-10\,GHz, MRC) SEDs in a statistically meaningful galaxy sample, the KINGFISH (Key Insights on Nearby Galaxies; a Far-Infrared Survey with Herschel, Kennicutt et al. 2011) with the 100-m Effelsberg telescope (PI: E. Schinnerer). The Effelsberg observations at 1.4\,GHz, 4.8\,GHz, 8.5\,GHz  combined with archive data at 10.5\,GHz (Niklas et al. 1997) allow us, for the first time, to determine the MRC bolometric luminosities and further present calibration relations for the MRC versus the monochromatic radio luminosities, and as a SFR tracer.  The radio SED is fitted using a Bayesian Marchov Chain Monte Carlo (MCMC) technique leading to measurements for the global nonthermal spectral index $\alpha_n$ and the thermal fraction (Tabatabaei et al. submitted). $\alpha_n$ is found to decrease with increasing the SFR surface density in galaxies, in agreement with our local studies. Hence, due to the star formation feedback, the CRE population is dominated by young and energetic particles in galaxies with high SFR.    


\begin{figure}
\center
\includegraphics[scale=0.5]{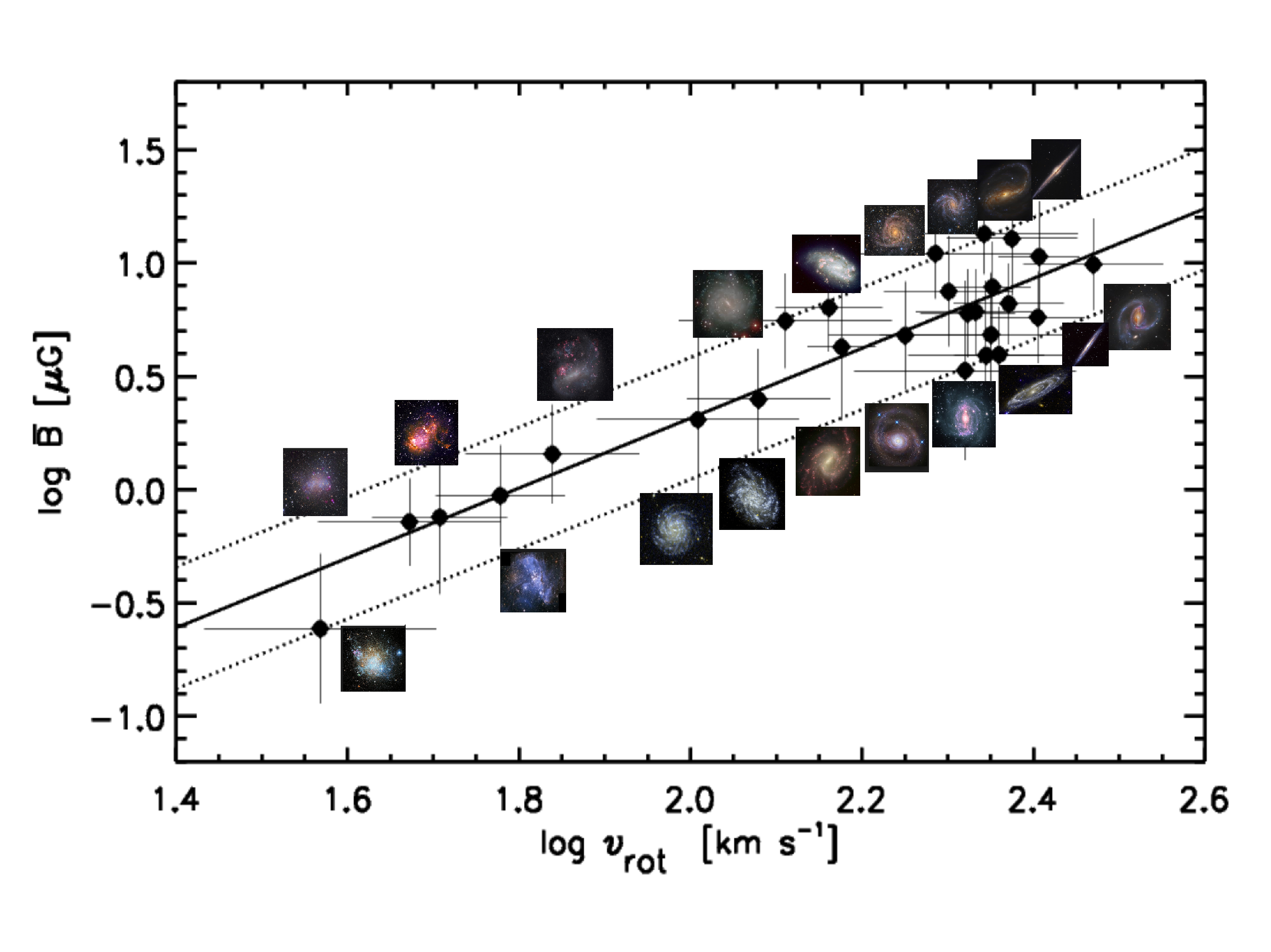} 
\caption{Strength of the large-scale magnetic field versus the rotation speed in nearby galaxies  (Tabatabaei et al. 2016). }
\end{figure}

%
%
\small  
%
\section*{Acknowledgments}   
%
FST acknowledges financial support from the Spanish Ministry of Economy and Competitiveness (MINECO) under grant number AYA2013-41243-P. 
%

%
\end{document}